\documentclass[doublecol]{epl2} 

\def\be{\begin{equation}}
\def\en{\end{equation}}    

\newcommand{\bi}[1]{\mbox{\boldmath$#1$}}
\newcommand{\av}[1]{\langle{#1}\rangle}
\def\p{\partial}
\def\bea{\begin{eqnarray}}
\def\ena{\end{eqnarray}}

\usepackage{graphicx}

\title{Structural and dynamical heterogeneities\\
in two-dimensional melting} 

\author{Hayato Shiba \and Akira  Onuki \and Takeaki Araki}
\shortauthor{H. Shiba, A. Onuki and T. Araki }

\institute{                    
Department of Physics, Kyoto University, 
Kyoto 606-8502, Japan
}
\pacs{64.70.D-}{Solid-liquid transitions}
\pacs{61.72.J-}{Point defects and defect clusters}
\pacs{83.10.Rs}{Computer simulation of molecular 
and particle dynamics}

\abstract{Using molecular dynamics simulation,   
we  study  structural and dynamical heterogeneities  
at  melting in  two-dimensional one-component systems 
with $36000$ particles. 
Between crystal and liquid we find 
intermediate hexatic states, 
where  the density  fluctuations 
are enhanced at small 
wave number $k$ as well 
as those of the six-fold orientational order parameter. 
Their structure factors both grow 
up to the smallest wave number equal to 
 the inverse system length. 
The intermediate scattering function
 of the density $S(k,t)$ is found to relax 
exponentially with  decay rate $\Gamma_k \propto k^z$ 
with $z \sim 2.6 $ at small $k$ in the hexatic phase.
}

\begin{document}

\maketitle

\section{Introduction}
Since a  simulation by 
Alder and Wainwright \cite{Alder}, 
much attention has  been paid to the 
two-dimensional (2D) melting 
in simple one component particle 
systems \cite{RMP}. However, it has been controversial 
whether the transition is first order 
as in three-dimensional melting  
\cite{Alder,RMP,Abraham,Toxvaerd,Weber}
or is continuous as predicted by 
 Halperin and Nelson \cite{HN}. 
They  presented a  
  defect-mediated  melting mechanism 
and a  ``hexatic phase''  
in a temperature (or density) window between crystal 
and liquid.  In the hexatic phase, the  
bond-orientation correlation function  $g_6(r)$ 
 of a sixfold orientation order  parameter $\psi_6({\bi r}) $ 
decays algebraically, indicating 
 a quasi-long-range  orientational correlation.
Afterwards their prediction 
has been confirmed in experiments 
\cite{Rice,Maret,Zahn,Quinn,Chaikin,Han} 
and in simulations 
\cite{Reichhardt,Rice1,Saito,Udink,Jaster,Binder,Watanabe,Mak}. 
In  the hexatic phase, defects have been observed to proliferate 
with increasing the temperature $T$ or decreasing the density $n$. 
As other theories,  Chui  proposed a   melting mechanism mediated by 
grain boundaries \cite{Chui},  while 
Saito argued that the 2D melting can be either  
continuous or first order depending on the specific details 
 of the system \cite{Saito}. 

As a marked feature, a number of 
experiments and simulations\cite{Alder,RMP,Abraham,Rice,Maret,Zahn,Quinn,Chaikin,Han,Reichhardt,Rice1,Harrowell}  
have  observed heterogeneities  in the hexagonal structures   
and in the particle trajectories, 
developing  around the transition. 
In particular, appreciable  dependence on the system size  
has been encountered in the  calculations 
of  the equation of state  \cite{Alder,Weber,Jaster,Mak}  
and the local  fluctuations 
of $\psi_6({\bi r})$\cite{Weber,Mak}. 
However,  the  heterogeneities   in the 2D melting 
 have  not yet been well understood.  
In this Letter,  we will  visualize them  
using  a disorder variable 
representing deviations from  the hexagonal order \cite{Hama}  
 and bond breakage used in analyzing 
glass dynamics \cite{yo}. 
It is also a fundamental issue  whether 
the isothermal compressibility $K_T=(\p n/\p p)_T/n$ 
remains finite or tends to infinity   in the hexatic phase. 
In simulations  \cite{Jaster,Mak}, 
the  pressure $p$ was  a weakly  decreasing function of 
$n$, apparently suggesting $(\p p/\p n)_T<0$,  
in the hexatic phase. 
Hence, 
we will  calculate the structure factor of the density 
$S(k)$ at small wave number $k$ 
to see whether the thermodynamic  
relation $\lim_{k\to 0}S(k)= n^2T K_T$ holds or not. 
The  intermediate scattering function $S(k,t)$ 
will then  emerge as a new informative 
quantity.

\section{Numerical Method}
Our 2D  system is composed of 
$N=36000$ particles interacting via 
a truncated Lennard-Jones potential of the form
\begin{equation} 
v(r) = 4\epsilon \left[( \sigma/r )^{12} - (\sigma/r )^6 \right] -C,
\end{equation}
which is characterized by the energy $\epsilon$ and the  range  
$\sigma$.   For  $r>r_{\rm cut}=3.2\sigma$, 
we set $v (r)=0$ with  the constant $C$  ensuring   
the continuity of $v(r)$ at the cut-off. 
The system volume $V$ is kept fixed  such that  
$\phi = N\sigma^2/V = 0.9$. Then the system length 
is given by $L =V^{1/2}=  200\sigma$. 
We integrated the  equations of motion 
using the St\"ormer-Verlet 
algorithm (a sort of the  leap-frog method) under the 
periodic boundary conditions using the 
Nos\'e-Hoover thermostat. 
The time step of integration is $0.002\tau$ with
\be 
\tau = \sigma \sqrt{m/\epsilon},
\en 
$m$ being the particle mass. 
In our  simulations,  we first quenched   the 
system from  $T=2\epsilon/k_B$   to $0.2\epsilon/k_B$ 
into a crystal state.  
After a relaxation time of $5\times 10^3\tau$, 
there was no appreciable time evolution in  physical 
quantities such as the pressure and  $g_6(r)$. 
We then  increased  $T$ to 
a desired value. The time $t$ is set equal to 
0 at this temperature increase. 
We continued  the simulation until $t=2\times 10^4\tau$.
Hereafter we will measure space, time, and $T$ 
in units of $\sigma$, $\tau$, and 
$\epsilon/k_B$, respectively.

In  2D dense particle systems, 
 a large fraction of 
the particles are enclosed by six particles 
and  the  local  order is  represented  by the  sixfold 
orientation \cite{HN}. 
We  define the orientation angle 
 $\alpha_j$ in the range $[-\pi /6, \pi/6]$   
 for each particle $j$ at position $\bm{r}_j$ using the complex number   
\begin{equation}
\Psi_j = \sum_{k\in\scriptsize{\textrm{bonded}}} \exp [6i\theta_{jk}] = |\Psi_j | e^{6i\alpha_j}, 
\end{equation}
where the summation is over particles ``bonded'' to the particle $j$. 
The two particles $j$ and $k$ are bonded if $|\bm{r}_j-\bm{r}_k| \le 1.25\sigma$.
$\theta_{jk}$ is the angle 
between $\bm{r}_j-\bm{r}_k$  and the $x$ axis \cite{Hama,yo}.  
Next we construct  another  nonnegative-definite 
 variable representing 
the degree of disorder  for each particle $j$ by
 \cite{Hama}    
\begin{equation}
D_j = 2\sum_{k\in\scriptsize{\textrm{bonded}}} [ 1-\cos 6(\alpha_j -\alpha_k)]. \end{equation}
Here $D_j$ is nearly zero for a perfect crystal, 
but  is large  in the range $5-20$ 
for  particles around defects.  Thus $D_j$ 
is  convenient in  visualizing  the structural 
inhomogeneity.

\begin{figure}
\begin{center}
\includegraphics[width=0.90\linewidth]{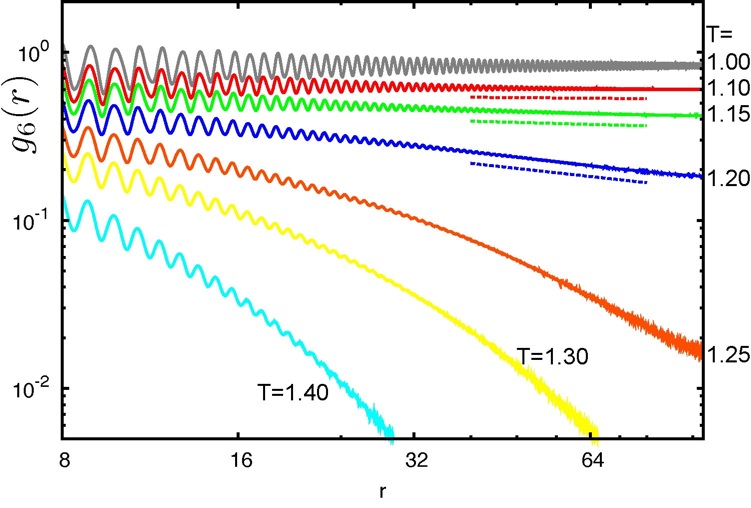}
\end{center}
\caption{
Bond-orientational correlation function $g_6(r)$
on a logarithmic scale for 
$T=1.0, 1.1, 1.15, 1.2, 1.25, 1.3,$ and $1.4$ from above,
obtained from data averaged over long-time and
over runs starting with 5 independent initial conditions.
Below the curves of $T=1.1, 1.15,$ and $1.2$,   
dotted line segments are guides to eye with slopes
$-0.03, -0.09,$ and $-0.36$, respectively.
}
\label{fig:fig1}
\end{figure} 

In terms of  $\alpha_j$ 
the  sixfold orientation order  parameter $\psi_6({\bi r}) $ 
is defined as \cite{HN}
\begin{equation}  
\psi_6 (\bm{r}) 
\equiv \sum_{j=1}^N e^{6i\alpha_j}\delta (\bm{r}-\bm{r}_j).
\end{equation}
In the hexatic phase, 
the bond-orientational correlation function 
decays algebraically as   
\begin{equation}
g_6(r)=\av{\psi_6({\bi r})\psi_6({\bi 0})^*} \sim r^{-\eta}, \label{eq:g6}
\end{equation}
where $r=|\bm{r}|$. Theoretically \cite{HN}, 
 the exponent $\eta$   depends  on $T$ 
and $n=N/V$ in the range $0<\eta<1/4$.

\begin{figure}
\begin{center}
\includegraphics[width=1.00\linewidth]{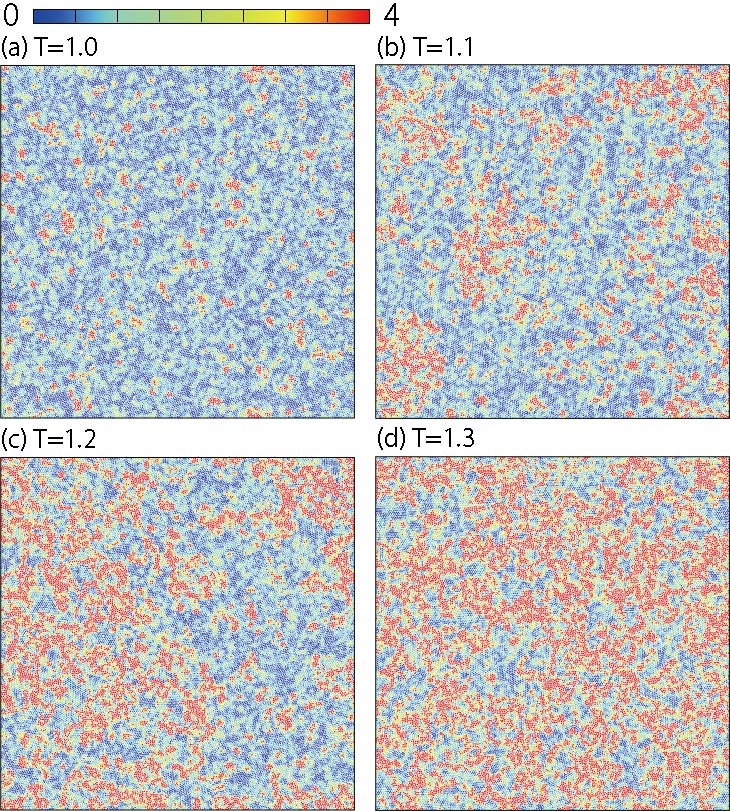}
\end{center}
\caption{Snapshots of disorder variable $D_j$ of  $36000$ particles 
at $t=1.2\times 10^4$ 
for (a) $T=1.0$ (crystal), (b) $T=1.1$ (hexatic), (c) $T=1.2$ (hexatic),
and (d) $T=1.3$ (liquid). 
Colors are given according to the  bar on the top. 
Particles with $D_j>4$ are written in red. 
}
\label{fig:fig2}
\end{figure}

\begin{figure*}
\begin{center}
\includegraphics[width=0.85\linewidth]{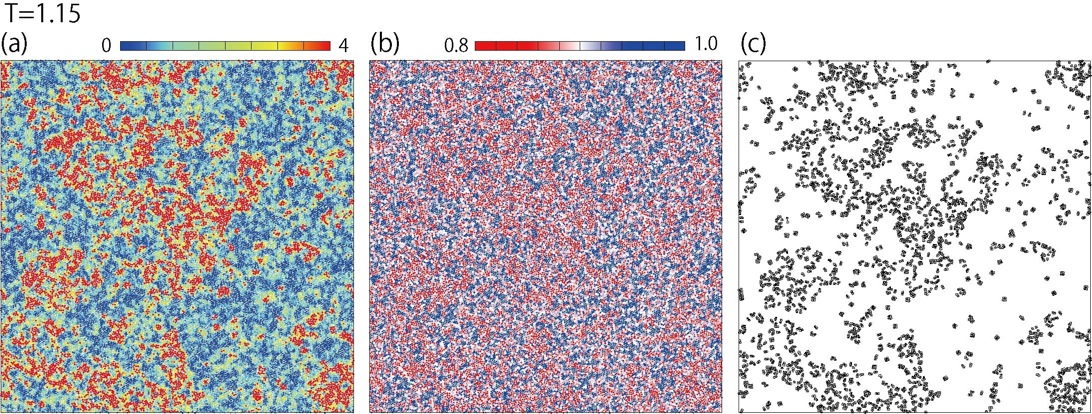}
\end{center}
\caption{
Snapshot of disorder variable $D_j$ (a),  
local packing density  $\rho_j$ (b), 
and particles with neighbor number 
different from six (c) 
for $36000$ particles at $T=1.15$. 
These three panels  exhibit heterogeneous patterns 
stemming from  the same  long-range 
structural disorder. 
}
\label{fig:fig3}
\end{figure*}

\section{Structural Heterogeneity}

\begin{figure}
\begin{center}
\includegraphics[width=0.85\linewidth]{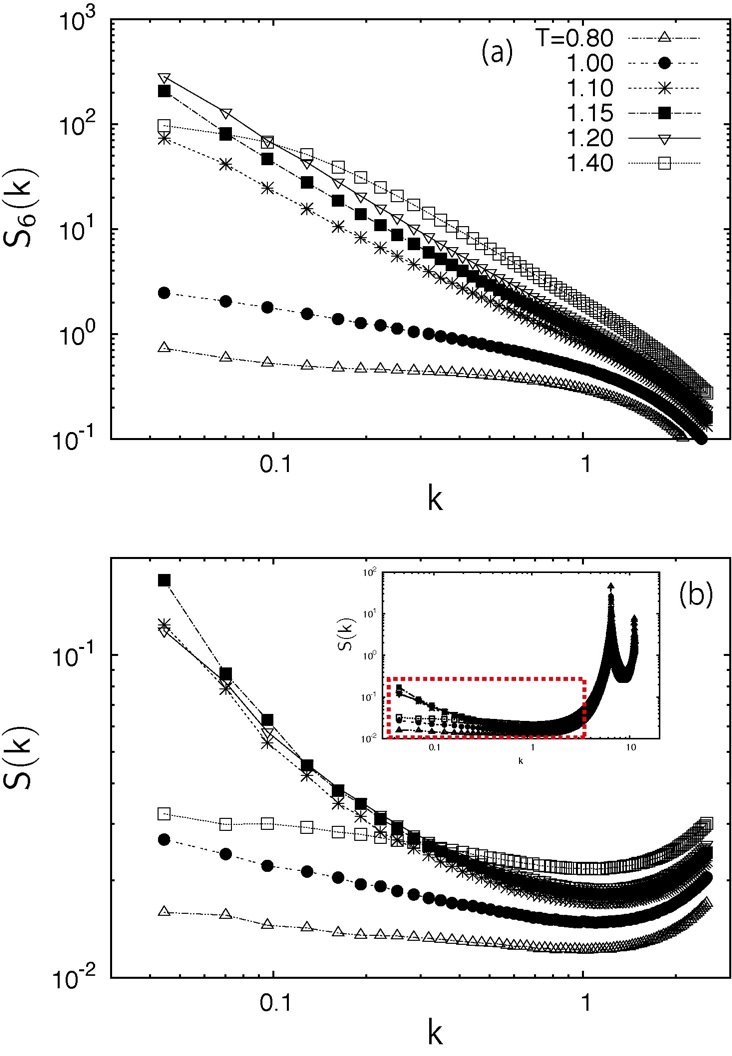}

\end{center}
\caption{Structure factors   
(a) $S_6(k)$ and (b) $S(k)$ for the six-fold 
orientation order  and  the density, respectively,   
for $T=0.8, 1.0, 1.1, 1.15, 1.2,$ and $1.4$, 
{calculated as  the long-time average.
Averages are taken also over 5 independent runs.
Use of mark in (b) is similar to (a).} 
They are enhanced at long wavelengths 
in the hexatic phase. 
}
\label{fig:fig4}
\end{figure}

In fig.~\ref{fig:fig1}, 
the curves of $g_6(r)$ are displayed, 
which  are the long-time averages 
over the snapshots  produced by $5$ independent runs 
in the range $10^4<t< 2\times 10^4$. 
In the hexatic phase at  $T=1.1$, $1.15$,  and $1.2$,   
 the exponent $\eta$  in eq. (\ref{eq:g6}) 
is $0.03, 0.09,$ and $0.36$, respectively, where 
0.03 at $T=1.1$ is very small and 
 0.36 at $T=1.2$ even  exceeds  the theoretical 
upper bound  $1/4$.  The hexatic-liquid and crystal-hexatic 
transitions are continuous, taking place at $T\cong 1.2$ and $1.0$ 
respectively.
For our system size, 
it is still difficult to determine 
the transition temperatures precisely.  
For $T\ge 1.25$  the system 
is in liquid with  $g_6(r)$ decaying exponentially, 
while for $T\le 1.0$ the system is in crystal 
without appreciable  decay of $g_6(r)$.

In fig.~\ref{fig:fig2}, 
we display snapshots of $D_j$ of all the particles 
in  a crystal phase at $T=1.0$ 
(a), in hexatic phases at $T=1.1$ (b) and 1.2 (c), 
and in a liquid phase at 
$T=1.3$ (d).  In fig.~\ref{fig:fig3} (a),  
a more expanded snapshot of $D_j$ 
is given at $T=1.15$. 
The average disorder parameter
$\overline{D}\equiv \sum_j D_j/N$ 
is 1.07, 1.72, 2.12, 2.55, and 3.32 for $T=1.0,1.1,1.15,1.2$ and $1.3$ respectively. 
In the hexatic phases ($1.1\le T\le 1.2$), 
  marked heterogeneity  emerges  
on large scales  among 
crystalline and disordered regions, though there are no 
sharp  boundaries.  The  patterns are fractal-like, 
resembling the critical fluctuations  
near the Ising criticality.  Also shown in 
fig.~\ref{fig:fig3} are (b) the local areal density $\rho_j$ 
(to be defined below) 
and (c) the particles with neighbor number being different from six. 
A common   particle configuration was used for  
 these three panels.   Here $v_j = \rho_j^{-1}$ is  the  volume
of particle $j$ in the Voronoi cell construction. 
We treat its  inverse  $\rho_j$  as the local density at 
the position of particle $j$. Its variance $\cal V$
($= \av{
\sum_j (\rho_j-\bar{\rho})^2/N}$ is about
  $0.05$ with ${\bar\rho}= \sum_j \rho_j/N\cong 0.9$ here.  
Comparing the two panels we 
can see that the particles with larger $D_j$ 
tend to have smaller $\rho_j$.
In the literature\cite{Rice,Quinn,Han}, defects have been  
detected around  the particles with 
five or seven neighbors, so we used  this Voronoi  method to 
produce fig.~\ref{fig:fig3} (c).  It  exhibits  
essentially the same  heterogeneity as that of $D_j$ (a),  
 though only discrete particles are selected. 
However, near melting, using $D_j$ 
is more quantitative and appropriate 
to characterize the diffuse 
disordered regions  
extending on large scales.

\begin{figure}
\begin{center}
\includegraphics[width=0.80\linewidth]{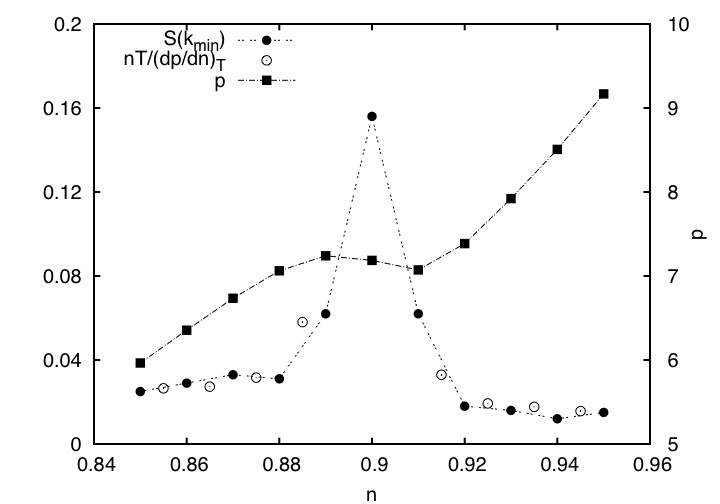} 
\end{center}
\caption{Pressure $p$ (right axis) 
vs density $n$ at $T=1.15$ with a small 
negative slope in the hexatic phase. 
Compared are resultant  $nT/({\partial p}/{\partial n})_T$ 
and $S(k_{\rm min})$ (left axis), which coincide 
outside the hexatic region 
confirming the the compressibility relation. 
In the hexatic phase, 
$(\p p/\p n)_T$ is apparently negative 
and the long wavelength limit $S(0)$ 
is not attained 
for our system size.  
}
\label{fig:fig5}
\end{figure}

The algebraic decay of $g_6(r)$ in eq.(4) arises from 
the heterogeneity in figs.~\ref{fig:fig2} and \ref{fig:fig3}. 
Furthermore, fig.~\ref{fig:fig3} (b) 
indicates that small density differences exist 
among the crystalline and disordered regions. 
In fig.~\ref{fig:fig4},  we thus show the structure factors 
of the hexagonal order 
and  the  density    
\begin{eqnarray}
&&S_6(k)  =  \int d\bm{r}\ e^{i\bm{k}\cdot\bm{r}}\ 
g_6(r)= \av{|\psi_{6\bm{k}}|^2},\\
&&S(k) 
 = \int d\bm{r}\ e^{i\bm{k}\cdot\bm{r}}\ 
\langle\delta\hat{n}(\bm{r}) \delta \hat{n}(\bm{0}) \rangle 
= \av{|\hat{n}_{\bm{k}}|^2},
\end{eqnarray}
where $\delta\hat{n}(\bm{r}) = 
\sum_{j} \delta (\bm{r} -\bm{r}_j)-n$ 
is the microscopic density deviation.  The 
$\psi_{6\bm{k}}$ and $\hat{n}_{\bm{k}}$ are the Fourier components 
of $\psi_6(\bm{r})$ and $\hat{n}(\bm{r})$.
These structure factors are the averages 
 over the angle of the wave vector 
${\bi k}= (k_x,k_y)$. 
The smallest wave number 
$k_{\rm min}$ is defined as 
$\pi (1+\sqrt{2})/L=0.035$,  
where $L=200$ is the system length.    
The structure factors 
at  $k= k_{\rm min}$ 
are  the averages of the data 
at ${\bi k}=  2\pi L^{-1} ( \pm 1, \pm 1)$ and 
$ 2\pi L^{-1} (\pm \sqrt{2}, \pm \sqrt{2})$.  
In fig.~\ref{fig:fig4}(a), the growth  
$S_6(k) \cong A_6k^{-2+\eta}$ can be seen 
at small $k$ in 
 the hexatic phase  at $T=1.1,1.15$, and $1.2$   
in accord with eq. (\ref{eq:g6}),  
where $A_6=0.53$ and $\eta=0.09$ at $T=1.15$. 
In fig.~\ref{fig:fig4} (b), $S(k)$ grows at small $k$ in the hexatic phase, 
but its amplitude  is very small and $S(k_{\rm min})$ 
remains smaller than the peak height 
at $k\sim 2\pi$ by two orders 
of magnitude (see the inset).   In fact, 
at $T=1.15$, its  curve may be fitted to 
$S(k)\cong 0.017+1.23\times 10^{-3}/ k^{1.55}$ 
for $k<1$.   The small coefficient 
($\sim 10^{-3}$) here  arises from 
small  density differences 
among the crystalline and disordered regions. 
For our system size, 
these structure factors  do not saturate 
even at  $k=k_{\rm min}$ in the hexatic phase.

If  $S(k)$ saturates to 
a long wavelength limit 
$S(0)=\lim_{k \to 0}S(k)$ 
in the thermodynamic limit $L\to \infty$ at fixed density, 
the compressibility is 
given by $K_T=  (\p n/\p p)_T/n=S(0)/n^2T$. 
From our simulation only, however,  
we cannot exclude 
the possibility of $S(k)\to \infty$ 
(as $k\to 0$)  
in the hexatic phase, where $(\p p/\p n)_T=0$ ultimately 
holds in the thermodynamic limit.  
As in fig.~\ref{fig:fig5}, we also performed simulations 
with  $36000$ particles by varying the volume $V$, where 
shown is the pressure $p$ (average of its  microscopic 
expression) vs the  density 
$n$ at $T=1.15$. 
Outside the hexatic region, 
the long wavelength limit 
$S(0)$ is attained 
and the compressibility relation 
$S(0)=nT/(\p p/\p n)_T$ surely  holds.  
Here, as in previous work \cite{Mak},  $p$  apparently 
exhibits a small negative slope 
in the hexatic density range. 
Thus   we need to use  much larger system sizes 
to settle  this issue. In such simulations, 
the long wavelength fluctuations (with $k<10^{-2}$) 
need to be equilibrated on extremely  long times. 

\begin{figure}
\begin{center}
\includegraphics[width=0.85\linewidth]{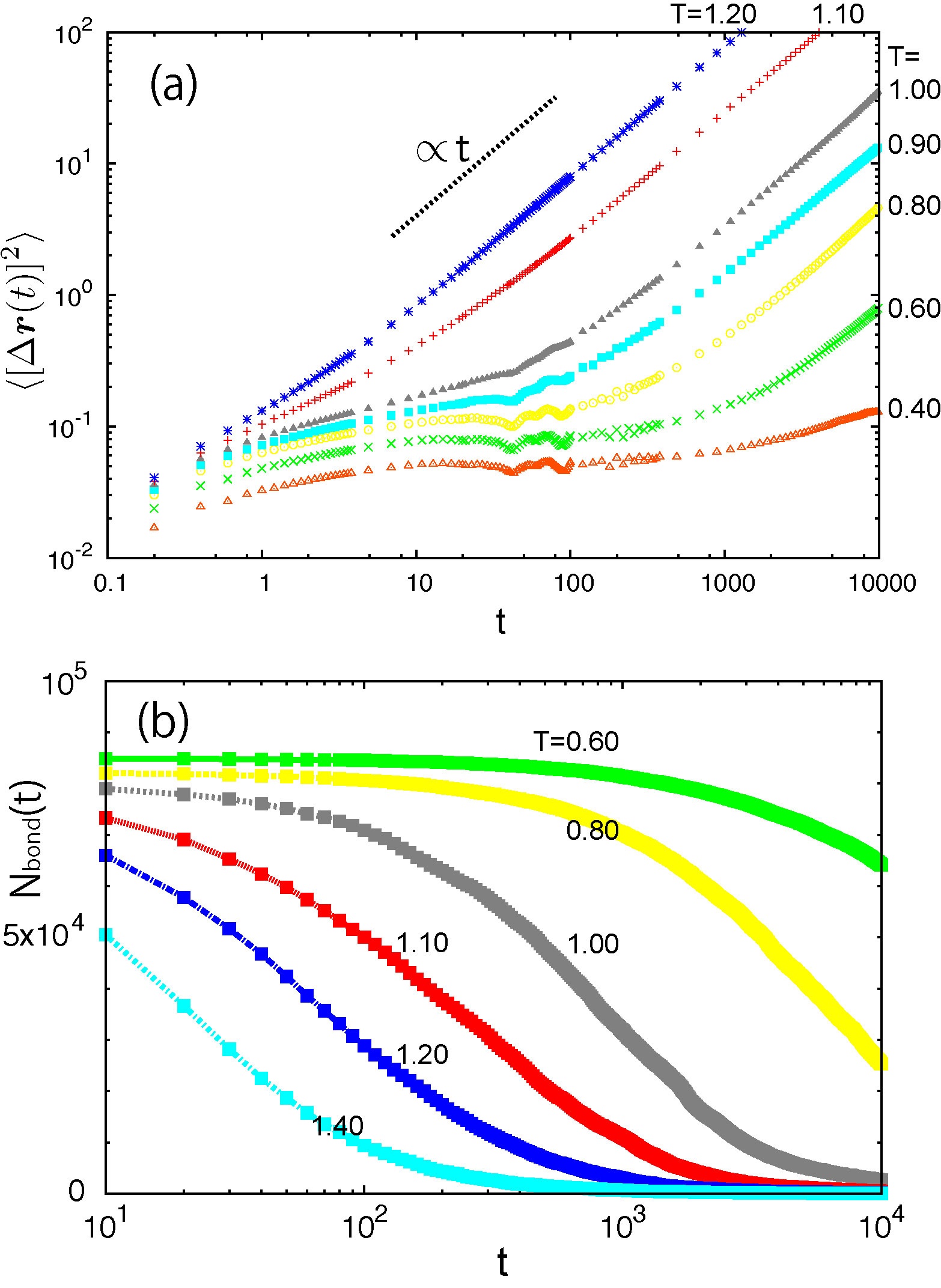} 
\end{center}
\caption{
(a) Mean-square displacement vs $t$ 
for various $T$ on a logarithmic scale, 
yielding  the diffusion constant $D_s$.
In crystal at  $T=0.6$ and $0.8$, its  
linear increase  after a plateau 
is due to  defect motion. 
(b) Surviving-bond 
number $N_b(t)$  vs $t$ on a semi-logarithmic  scale,  
decaying as $e^{-t/\tau_b}$. Here $D_s \sim 1/\tau_b$. 
}
\label{fig:fig6}
\end{figure}

\begin{figure*}
\begin{center}
\includegraphics[width=0.9\linewidth]{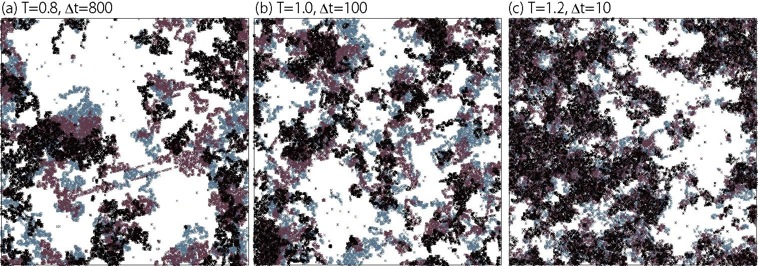}
\end{center}
\caption{
Snapshots of broken bonds in three consecutive time 
intervals $[t_0+ (\ell-1)\Delta t, t_0+\ell \Delta t]$ ($\ell=1,2,3)$ 
for (a)$\Delta t=800$ at $T=0.8$, 
(b)$\Delta t=100$ at $T=1.0$, 
and (c) $\Delta t=10$  at $T=1.2$. Colors: 
ultramarine ($\ell=1$), magenta ($\ell=2$), 
and black ($\ell=3$) in the chronological  order.  Corresponding 
 snapshots  of $D_j$   of (a) and (b) at $\ell=3$ 
are shown in fig.~\ref{fig:fig2} (b) and (c) 
}
\label{fig:fig7}
\end{figure*}

\section{Dynamics}


The   dynamics has not yet been well 
studied at  the 2D  melting. 
For the particle displacement  $\Delta \bm{r}_j (t) = 
\bm{r}_j (t+t_0) - \bm{r}_j (t_0)$ in time interval $t$,
 fig.~\ref{fig:fig6} (a) displays  
 the mean square displacement, 
$
\langle [\Delta \bm{r}(t)]^2\rangle = 
\sum_{j} \langle [\Delta \bm{r}_j(t)]^2 \rangle/N,
$
which is the average over all the particles and 
over the simulation time. 
The linear behavior 
$\langle [\Delta \bm{r}(t)]^2\rangle \cong 4D_st$ 
can be seen at long times   for $T\ge 0.6$. 
The diffusion constant $D_s$  thus obtained increases as   
$0.181$, $1.13$, $3.32$, $8.23$, $60.5$, 
and $188$ for $T=0.6,0.8,0.9, 1, 1.1, $ and 1.2, 
respectively, in units of $10^{-4}
\sigma^{-2}\tau^{-1}$. 
In   crystal ($0.6\le T \le 1.0$), the curves 
exhibit a plateau 
followed by the linear growth. 
Similar two-step behavior is well-known in 
supercooled liquids \cite{yo}, 
but it is here due to  motions of defect clusters  
  composed  of several particles 
with finite $D_j$ (see fig.~\ref{fig:fig2} (a) ). 
Such clusters  were  observed in 2D colloidal systems \cite{Han}.  
In the hexatic phase,  on the other hand, the plateau disappears  
and $D_s$ grows abruptly.

 To examine the particle-configuration 
changes,  we introduce the bond breakage   
\cite{yo}. For each particle 
 configuration  at a time $t_0 (\sim 10^4)$ 
after long annealing, 
a pair of particles $i$ and $j$ is considered to be bonded if
\begin{equation}
r_{ij}(t_0) = |\bm{r}_i(t_0)-\bm{r}_j (t_0)| \le A_1, 
\end{equation}
where we set $A_1=1.2$ 
(around the peak distance of the pair correlation function). 
 After a time interval $ t$, 
the bond is regarded
to be broken if 
\begin{equation}
r_{ij}(t_0 + t)\ge A_2,
\end{equation}
where we set 
$A_2=1.6$. In fig.~\ref{fig:fig6} (b), we plot 
the number of the surviving  bonds $N_b(t)$ 
vs $t$  for various $T$. It  is 
equal to the initial bond number $N_b(0)$ 
($\sim 8.4\times 10^4$)  minus the 
number of the broken bonds.   It may fairly be fitted 
to the exponential  form $e^{-t/\tau_b}$. 
The bond life time $\tau_b$  
is determined by $N_b(\tau_b)=N_b(0)/e$. 
Then  $\tau_b=91$,  $8.1$, $1$, 
$0.3$, and $0.09$ 
for  $T=0.6,0.8, 1, $ and 1.2, 
respectively, in units of $10^{3}\tau$. 
Here we notice  that the product 
$D_s\tau_b$ is between 1 and 2, where $D_s$ is the diffusion 
constant determined in fig.~\ref{fig:fig6} (a).  
Thus 
\begin{equation} 
D_s\sim \tau_b^{-1}, 
\end{equation}
which demonstrates that  
 the particle motions are 
caused by the configuration changes in our 
jammed states.

We examine   how 
 the  dynamic heterogeneity evolves  in time. 
In fig.~\ref{fig:fig7}, we pick up 
the particles with   broken bonds 
in three consecutive time intervals 
and mark them 
in ultramarine, magenta, and black in this order. 
Here we set  (a)$\Delta t =800$ at $T=0.8$, 
(b)$\Delta t =100$  at $T=1.0$,   and 
(c)$\Delta t =10$  at  $T=1.2$.   
In  crystal (a), the evolution is 
due to the motions of defect clusters taking place 
in the form of string-like trajectories in each event. 
Such trajectories accumulate to 
form large-scale dynamic heterogeneity 
on long time scales, though the 
defect number  is small at each time in crystal. This 
picture explains  the two-step behavior of 
$\langle [\Delta \bm{r}(t)]^2\rangle$ 
in crystal in fig.~\ref{fig:fig6} (a). 
In addition, two straight lines arising from    dislocation gliding 
can be seen in (a), but such slip motions  are rare in our system. 
In (b), the system is still in crystal, but 
the time scale of bond breakage 
is much faster.  In the hexatic phase (c), 
the particles in the disordered  regions 
 are relatively  mobile, while those in the crystalline regions 
are nearly immobile. The mobility of the particles  is 
strongly correlated to the disorder variable  $D_j$ 
visualized in fig.~\ref{fig:fig2} (c).
The dynamic heterogeneity 
can be seen over a rather broad  temperature range 
around the melting, where the time scale changes 
dramatically 
\cite{Zahn,Han,Reichhardt,Rice1,Harrowell,Hama}.

\begin{figure}
\begin{center}
\includegraphics[width=0.85\linewidth]{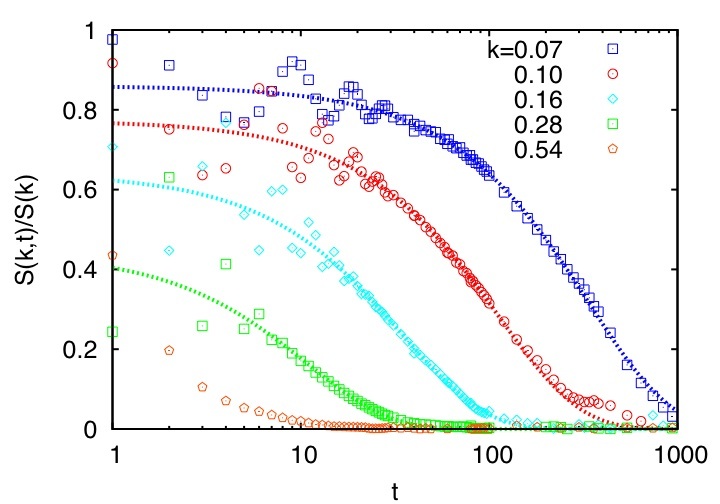} 
\end{center}
\caption{Normalized 
intermediate scattering function $S(k,t)/S(k)$ in the hexatic 
phase at $T=1.15$ for various small $k$. 
After the initial acoustic damped  oscillation, it can be 
fitted to the exponential decay of the form  $A_k e^{-\Gamma_k t}$  
(solid lines) with 
$\Gamma_k \propto k^{2.6}$. 
}
\label{fig:fig8}
\end{figure}

Since the large-scale density fluctuations 
are enhanced as in fig.~\ref{fig:fig4} (b), 
we are interested in their relaxation. 
As shown in fig.~\ref{fig:fig8}, we calculated the 
total intermediate scattering function
\begin{equation}
S(k,t) = \int d\bm{r}\ e^{i\bm{k}\cdot\bm{r}}\ \langle \delta\hat{n} (\bm{r},t) \delta\hat{n} (\bm{0},0 )\rangle
\end{equation}
 in the hexatic phase at 
$T=1.15$ where the initial value 
$S(k)$ is enhanced as in fig.~\ref{fig:fig4} (b). 
In an early stage  it undergoes an oscillatory 
decay arising from the acoustic propagation. 
We find that it  then decays exponentially as 
\begin{equation}
S(k,t)\cong  S(k) A_k e^{-\Gamma_k t}, 
\end{equation}
for  small $k \ll 1$. 
The amplitude $A_k$ approaches unity 
and the decay rate behaves as 
$\Gamma_k \sim k^{z}$ with $z\sim 2.6$ for small $k$. 
The dotted lines in  fig.~\ref{fig:fig8} 
represent  this exponential form, which are excellently fitted 
to the numerical data.  This slow decay arises from 
the evolution of large-scale density  heterogeneities  
produced by the structural  fluctuations.
In the hexatic phase, the 
diffusion constant 
depends on the wave number as $D_k \propto k^{z-2}$ 
if it is introduced by $D_k=\Gamma_k/k^2$. 
This is analogous to the thermal diffusion constant 
at the gas-liquid criticality, where $z=3$ 
in three dimensions \cite{Onukibook}.  
However, we cannot explain 
this exponential  relaxation in the hexatic phase at present.  
In passing, we  also calculated $S(k,t)$ in crystal and liquid 
outside the hexatic temperature window, where 
 the  density fluctuations are much suppressed at small $k$. 
There, the  long wavelength   
relaxation of $S(k,t)$ 
is due to hydrodynamic thermal 
diffusion with $\Gamma_k \propto k^2$ for small $k$.

\section{Summary and Comments}

We have examined  the heterogeneities  
in structure and dynamics at two-dimensional melting  
of one-component systems at fixed volume $V$.  In terms of 
 the disorder variable $D_j$, 
structurally  heterogeneous patterns 
have emerged  unambiguously  
among the crystalline and disordered regions 
in the hexatic phase in figs. 2 and 3.  
Though very weak, we have  
noticed  the presence of large-scale  density fluctuations.   
As a result, at small wave number $k$, 
the structure factor $S_6(k)$ of the sixfold orientation  order 
grows strongly as in fig.~\ref{fig:fig4} (a), while the structure factor 
$S(k)$ of the density grows similarly but its 
 amplitude is very small as 
in fig.~\ref{fig:fig4} (b).  Dynamically  heterogeneous 
patterns have been   obtained 
in fig.~\ref{fig:fig7} in the crystal and hexatic phases.
Crystal states are dynamically 
heterogeneous on long time scales 
due to relatively rapid motions of 
 defect clusters.    In the hexatic phase, 
dynamics  is  heterogeneous on  short time scales, 
where  the  particles with high $D_j$ tend to be 
mobile than those with small $D_j$. We have also 
calculated the intermediate scattering function $S(k,t)$, 
which has turned out to relax  
 exponentially with the dynamic exponent 
$z$ about  ${2.6}$ at small $k$ 
in the hexatic phase.

We make further  comments. (i) In our simulation, mesoscopic 
coexistence of the ordered and disordered regions 
is realized dynamically  
in  the hexatic phase.  
There are  no sharp  boundaries 
between the two regions and the free energy penalty 
due to the structural inhomogeneity should 
be very small.  These structural  fluctuations 
resemble the critical fluctuations 
in Ising systems. It remains puzzling 
whether or not the long wavelength limit 
of $S(k)$ tends to a finite constant  in the 
thermodynamic limit. If so, 
it follows a finite compressibility. 
(ii) Growth (shrinkage) of the disordered regions 
gives rise to  an  increase (a decrease)  in 
the pressure $p$  at fixed volume. 
We should perform constant-pressure simulations also to examine 
whether or not the hexatic phase exists depends 
on the boundary condition. 
(iii) In binary  mixtures  the size dispersity 
serves to pin the particle motions and 
the relaxation times  are much longer  than 
in one-component systems. 
(iv)   Shear flow has been 
applied to glassy and   
polycrystal systems \cite{yo,Hama1,Onukibook}. It is intriguing how 
applying shear can  affect  the hexatic state and 
the scenario of the 2D melting.

\begin{acknowledgments}
We thank  T. Hamanaka, N. Ito, T. Uneyama, R. Yamamoto, 
 S. Yukawa, H. Watanabe, R. Okamoto, and T. Kawasaki for valuable  discussions.
This work was supported in part by Grant-in-Aid for Scientific Research
on the Priority Area ``Soft Matter Physics'' 
and for the Global COE Program ``The Next Generation of Physics, Spun from Universality and Emergence''
from the MEXT of Japan.
H. S. was supported by JSPS.
\end{acknowledgments}




\end{document}